\documentclass[conference,a4paper]{IEEEtran}
\usepackage{xcolor}
\usepackage{balance}
\usepackage{cite}
\usepackage{multirow}
\usepackage[pdftex]{graphicx}
\usepackage{amsmath,amssymb,amsfonts}
\usepackage{textcomp}
\usepackage[caption=false,font=footnotesize]{subfig}
\usepackage[nolist]{acronym}
\usepackage[capitalize]{cleveref}
\usepackage{nicefrac}
\usepackage{graphicx}  % remove 'demo' option for your real document
\crefname{table}{Tab.}{Tabs.}
\Crefname{table}{Tab.}{Tabs.}
\crefformat{equation}{(#2#1#3)}

\newcommand\norm[1]{\left\lVert#1\right\rVert}

\newcommand{\NormalBrackets}[1]{ \left( #1 \right) }

\begin{document}
\begin{acronym}
\acro{AWG}[AWG]{Arbitrary Waveform Generator}
\acro{CW}[CW]{Continuous Wave}
\acro{ADC}[ADC]{analog-to-digital converter}
\acro{DAC}[DAC]{digital-to-analog converter}
\acro{HST}[HST]{High-Speed Train}
\acro{IF}[IF]{Intermediate Frequency}
\acro{LO}[LO]{local oscillator}
\acro{LSF}[LSF]{local scattering function}
\acro{MMW}[mmWave]{millimeter wave}
\acro{OFDM}[OFDM]{orthogonal frequency-division multiplexing}
\acro{PCB}[PCB]{Printed Circuit Board}
\acro{SMD}[SMD]{Surface Mount Device}
\acro{SNR}[SNR]{signal-to-noise ratio}
\acro{SINR}[SINR]{signal-to-interference-and-noise ratio}
\acro{RF}[RF]{radio frequency}
\acro{V2X}[V2X]{vehicle-to-everything}
\acro{IC}[IC]{Integrated Circuit}
\acro{FPGA}[FPGA]{Field Programmable Gate Array}
\acro{ISI}[ISI]{Inter-Symbol Interference}
\acro{OOBA-MRC}[OOBA-MRC]{out-of-band aided maximal ratio combining}
\acro{CS}[CS]{compressed sensing}
\acro{CSIT}{channel state information at the transmitter}
\acro{CSI}{channel state information}
\acro{ML}{machine learning}
\acro{IL}{insertion loss}
\acro{RNN}{recurrent neural networks}
\acro{EVM}{error vector magnitude}
\acro{OMP}{orthogonal matching pursuit}
\acro{PA}[PA]{power amplifier}
\acro{LNA}[LNA]{low noise amplifier}
\acro{EIRP}{effective isotropic radiated power}
\acro{LSTM}{long short-term memory}
\acro{GRU}{Gated Recurrent Unit}
\acro{FDD}{frequency-division duplex}
\acro{TDD}{time-division duplex}
\acro{CIR}{channel impulse response}
\acro{CTF}{channel transfer function}
\acro{PDP}{power delay profile}
\acro{DSD}{Doppler power spectral density}
\acro{IFFT}{Inverse Fast Fourier Transform}
\acro{ITS}[ITS]{intelligent transportation systems}
\acro{5G}[5G]{fifth generation}
\acro{NR}[NR]{new radio}
\acro{QAM}[QAM]{quadrature amplitude modulation}
\acro{ICI}[ICI]{Inter-Carrier Interference}
\acro{MSE}[MSE]{mean squared error}
\acro{BER}[BER]{bit error ratio}
\acro{RMS}[RMS]{root-mean-square}
\acro{TDL-A}[TDL-A]{tapped delay line A}
\acro{TDL-D}[TDL-D]{tapped delay line D}
\acro{ISI}[ISI]{Inter-Symbol Interference}
\acro{DFT}[DFT]{discrete Fourier transform}
\acro{IDFT}[IDFT]{inverse discrete Fourier transform}
\acro{CTF}[CTF]{channel transfer function}
\acro{DR}[DR]{dynamic range}
\acro{HPBW}[HPBW]{half-power beamwidth}
\acro{DPSS}[DPSS]{discrete prolate spheroidal sequences}
\acro{CDF}[CDF]{cumulative distribution function}
\acro{URLLC}[URLLC]{ultra-reliable low-latency communication}
\acro{3GPP}[3GPP]{3rd Generation Partnership Project}
\acro{MIMO}[MIMO]{multiple-input multiple-output}
\acro{SISO}[SISO]{single-input single-output}
\acro{MRC}[MRC]{maximal ratio combining}
\acro{AWGN}[AWGN]{additive white Gaussian noise}
\acro{LS}[LS]{least-squares}
\acro{SVD}[SVD]{singular value decomposition}
\acro{SE}[SE]{spectral efficiency}
\acro{ULA}[ULA]{uniform linear array}
\acro{FSPL}[FSPL]{free space path loss}
\acro{LOS}[LOS]{line-of-sight}
\acro{NLOS}[NLOS]{non-line-of-sight}
\acro{AoA}[AoA]{angle-of-arrival}
\acro{MUSIC}[MUSIC]{multiple signal classification}
\acro{EVD}[EVD]{eigenvalue decomposition}
\acro{AoD}[AoD]{angle-of-departure}
\acro{RMSE}[RMSE]{root mean squared error}
\acro{SIMO}[SIMO]{single-input multiple-output}
\acro{MAC}[MAC]{medium access control}
\acro{eMBB}[eMBB]{enhanced mobile broadband}
\acro{SVM}[SVM]{support vector machine}
\acro{VNA}[VNA]{vector network analyzer}
\acro{CNN}[CNN]{convolutional neural network}
\acro{DNN}[DNN]{deep neural network}
\acro{EE}[EE]{energy efficiency}
\acro{ReLU}[ReLU]{rectified linear unit}
\acro{ZP}[ZP]{zero padding}
\acro{BN}[BN]{batch normalization}
\acro{NN}[NN]{neural network}
\acro{ADAM}[ADAM]{adaptive moment estimation}
\acro{AQNM}[AQNM]{additive quantization noise model}
\end{acronym}
%\title{Digital Beamforming for mmWave MIMO with Low-Resolution Converters}
%\title{Channel Estimation for Digital Beamforming mmWave MIMO with Low-Resolution Converters}
%\title{Channel Estimation for Digital Beamforming mmWave MIMO with Low-Resolution DACs/ADCs}
%\title{Digital Beamforming for Millimeter Wave MIMO with Low-Resolution DACs/ADCs: A Comparison of Channel Estimation Methods}
\title{Performance Analysis of Digital Beamforming mmWave MIMO with Low-Resolution DACs/ADCs}
% A Comparison of Hybrid Beamforming and Digital Beamforming With Low-Resolution ADCs for Multiple Users and Imperfect CSI

\author{\IEEEauthorblockN{
Faruk Pasic\IEEEauthorrefmark{1}\thanks{The work of F.~Pasic has been supported by TU Wien with a KuWi Scholarship.
The work of M.~Mussbah has been funded by the Christian Doppler Laboratory for Digital Twin assisted AI for sustainable Radio Access Networks, Institute of Telecommunications, TU Wien.},
Mariam Mussbah\IEEEauthorrefmark{1}\IEEEauthorrefmark{2},
Stefan Schwarz\IEEEauthorrefmark{1},
Markus Rupp\IEEEauthorrefmark{1},
Fredrik Tufvesson\IEEEauthorrefmark{3}, \\ and
Christoph F. Mecklenbräuker\IEEEauthorrefmark{1}
}%

\IEEEauthorblockA{\IEEEauthorrefmark{1}% 2nd affiliations
Institute of Telecommunications, TU Wien, Vienna, Austria}
\IEEEauthorblockA{\IEEEauthorrefmark{2}% 2nd affiliations
Christian Doppler Laboratory for Digital Twin assisted AI for sustainable Radio Access Networks}
\IEEEauthorblockA{\IEEEauthorrefmark{3}% 3rd affiliations
Department of Electrical and Information Technology, Lund University, Lund, Sweden}
\IEEEauthorblockA{faruk.pasic@tuwien.ac.at}
}

\IEEEoverridecommandlockouts 

\makeatletter
\def\thanks#1{\protected@xdef\@thanks{\@thanks
        \protect\footnotetext{#1}}}
\makeatother

\maketitle

\begin{abstract}
Future wireless communications will rely on \ac{MIMO} beamforming operating at \ac{MMW} frequency bands to deliver high data rates.
To support flexible spatial processing and meet the demands of latency-critical applications, it is essential to use fully digital \ac{MMW} \ac{MIMO} beamforming, which relies on accurate channel estimation.
However, ensuring power efficiency in fully digital \ac{MMW} \ac{MIMO} systems requires the use of low-resolution \acp{DAC} and \acp{ADC}. %must be operated at a low quantization resolution to stay within reasonable power levels.
The reduced resolution of these quantizers introduces distortion in both transmitted and received signals, ultimately degrading system performance.
%The limited resolution of these components introduces quantization noise, which distorts both the transmitted and received signals, leading to performance degradation. 
In this paper, we investigate the channel estimation performance of \ac{MMW} \ac{MIMO} systems employing fully digital beamforming with low-resolution quantization, under practical system constraints.
We evaluate the system performance in terms of \ac{SE} and \ac{EE}.  
Simulation results demonstrate that a moderate quantization resolutions of 4-bit per \ac{DAC}/\ac{ADC} offers a favorable trade-off between energy consumption and achievable data rate.  
\end{abstract}
\vskip0.5\baselineskip
\begin{IEEEkeywords}
mmWave, MIMO, channel estimation, digital beamforming, quantization.
\end{IEEEkeywords}

\acresetall

\section{Introduction}
\Ac{MMW} communication systems combined with \ac{MIMO} techniques offer the potential to meet growing data rate demands~\cite{Molisch2025, Pasic2023_mag}.
To fully exploit the spatial multiplexing and beamforming capabilities of \ac{MIMO}, fully digital beamforming architectures are required.
These architectures enable simultaneous channel estimation in multiple spatial directions within a single time interval, significantly reducing the time overhead for link establishment and making them particularly well-suited for latency-critical applications~\cite{Liu2020}.

Nevertheless, compared to power-efficient analog or hybrid beamforming architectures, fully digital beamforming in wideband \ac{MMW} systems tends to be significantly more power-intensive.
This is primarily due to the fact that each \ac{RF} chain requires a dedicated pair of \acp{DAC} at the transmitter and \acp{ADC} at the receiver.
To reduce power consumption, fully digital beamformers must utilize low-resolution converters~\cite{Singh2009}, which in turn introduce distortion in both the transmitted and received signals.
Therefore, it is essential to evaluate the performance of such systems under practical hardware constraints.

The performance of digital \ac{MMW} beamforming systems employing low-resolution quantizers has been explored to some extent in the literature~\cite{Ribeiro2018, Dutta2020, Roth2018}. 
However, these studies are limited in terms of the range of antenna configurations and/or the channel conditions considered. 
Moreover, to the best of the author's knowledge, the effect of low-resolution quantization on advanced \ac{MMW} \ac{MIMO} channel estimation techniques remains unexplored.

\textbf{Contribution:}
In this paper, we evaluate the performance of fully digital low-resolution \ac{MMW} \ac{MIMO} systems for various channel estimation methods.
Our analysis considers different antenna configurations and varying channel conditions characterized by the Rician $K$-factor.
We investigate the impact of reduced quantization resolution through simulations in terms of \ac{SE} and \ac{EE}.

\textbf{Organization:}
\cref{sec:system_model} describes the system model along with the adopted quantization noise model.
A simulation-based performance evaluation is presented in~\cref{sec:comparison}, and \cref{sec:conclusion} concludes the paper.

\textbf{Notation:} 
We use $x$ to denote scalars, ${\mathbf x}$ for vectors (bold lowercase), and ${\mathbf X}$ for matrices (bold uppercase).
The $j$-th column of a matrix ${\mathbf X}$ is denoted by $\mathbf{X}_{:, j}$.
Transpose and Hermitian transpose operations are denoted by $\left( \cdot \right) ^{\rm T}$ and $\left( \cdot \right) ^{\rm H}$, respectively.
The Euclidean norm is written as $\lVert \cdot \rVert$, the Frobenius norm as $\lVert \cdot \rVert_F$ and the expectation operator as $\mathbb{E} \NormalBrackets{ \cdot }$.

\section{System Model} \label{sec:system_model}
We study a point-to-point \ac{MMW} \ac{MIMO} system operating in the radiative far-field, assuming perfect synchronization between the transmitter and receiver.
The transmitter is equipped with  $M_{\rm Tx}$ and the receiver with $M_{\rm Rx}$ antenna elements.
Each antenna is equipped with its own \ac{RF} chain, enabling fully digital beamforming.
The system employs \ac{TDD} to ensure channel reciprocity.
Additionally, the system utilizes \ac{OFDM} and quadrature amplitude modulation with $N$ subcarriers.

At the $n$-th \ac{OFDM} subcarrier, the channel matrix is denoted as $\mathbf{H} [n] \in \mathbb{C}^{M_{\rm Rx}  \times M_{\rm Tx} }$, based on the equivalent complex baseband representation.
The channel is modeled as a frequency-selective Rician fading channel based on the \ac{3GPP} \ac{MIMO} channel model, with an adjustable Rician $K$-factor~\cite{3gpp.38.901}.

\subsection{Link Establishment} \label{subsec:link_establishment}
Link establishment consists of two phases: training and data transmission. 
During the training phase, the wireless channel is estimated and the resulting channel estimate is used in the subsequent data transmission phase.

% Transmitter impairments
% The samples of ηu [n] from different users u and time instances n are independent.
% Including the transmit power Pu , this is the classical Error Vector Magnitude (EVM) definition only considering transmitter
% impairments [23]. 
% In this work we only consider the transmitter EVM. 
% Thus, the noise here combines standard thermal noise and non-linear contributions from the whole transmitter hardware,
% including Digital-to-Analog-Converter (DAC), PAand the phase
% noise of the Local Oscillators (LO).

\subsubsection{Training Phase} \label{subsubsec:training_phase}
In the training phase, known pilot symbols are transmitted. 
The input-output relationship is given by
\begin{equation} 
    \mathbf{y} [n] = 
    \mathbf{H} [n] 
    \NormalBrackets{\boldsymbol{\phi} [n] + \mathbf{d} [n] } 
    + \mathbf{w} [n],
    \label{eq:training_input_output}
\end{equation}
where $\mathbf{y} [n] \in \mathbb{C}^{M_{\rm Rx} \times 1} $ represents the unquantized received symbols, $\boldsymbol{\phi} [n] \in  \mathbb{C}^{M_{\rm Tx} \times 1}$ represents the transmitted pilot symbols, $\mathbf{d} [n] \in \mathcal{CN}(0,\sigma_{\rm d}^2 \mathbf{I}_{M_{\rm Tx}})$ models impairments from the entire transmit chain as in~\cite{Gupta2012}, and $\mathbf{w} [n] \in \mathcal{CN}(0,\sigma_{w}^2 \mathbf{I}_{M_{\rm Rx}})$ is the \ac{AWGN}.
The pilot symbols for the $t$-th transmit antenna are defined as
\begin{equation}
    \mathrm{\phi}_t [n] = 
    \begin{cases}
        \mathrm{\phi} [n], \, n \in \{ t, t+M_{\rm Tx}, \ldots, N - M_{\rm Tx} + t \}
        \\
        0, \quad \, \, \, \text{else}
    \end{cases}
     \label{eq:pilot_allocation}
\end{equation}
where $\mathrm{\phi} [n]$ are one-dimensional complex symbols drawn from a \ac{QAM} alphabet.
The transmitter impairments include \ac{RF} impairments (such as those from the \ac{PA} and \ac{LO} phase noise) with a combined power denoted by $\sigma_{\rm RF}^2$, as well as quantization noise from \ac{DAC} with power $\sigma_{\rm q, Tx}^2$~\cite{Gupta2012}.
To model the effect of low-resolution quantization, we adopt the \ac{AQNM}, following the approach used in~\cite{Dutta2020, Fletcher2007}.
Adopting this model, the power of the quantization noise from the \ac{DAC} is given by
\begin{equation}
    \sigma_{\rm q, Tx}^2 =    
    \frac{1}{M_{\rm Tx}} 
    \Upsilon\NormalBrackets{n_{\rm b}} 
    \NormalBrackets{ 1 - \Upsilon\NormalBrackets{n_{\rm b}}}, 
\end{equation}
with the inverse coding gain approximated by $\Upsilon\NormalBrackets{n_{\rm b}} \approx \frac{\pi \sqrt{3}}{2} n_{\rm b}^{-2}$~\cite{Orhan2015},
where $n_{\rm b}$ denotes the number of quantization bits and $\Upsilon\NormalBrackets{n_{\rm b}} = 0$ corresponds to the ideal case of infinite quantization resolution.
Therefore, the total impairment power is given by $\sigma_{\rm d}^2 = \sigma_{\rm RF}^2 + \sigma_{\rm q, Tx}^2$.    
Since both $\mathbf{d} [n]$ and $\mathbf{w} [n]$ are mutually independent and  Gaussian, they can be combined into a single noise term, we can rewrite~\cref{eq:training_input_output} as
\begin{equation}
    \mathbf{y} [n] = 
    \mathbf{H} [n] \boldsymbol{\phi} [n] + \mathbf{w}' [n],
%    \underbrace{\mathbf{H} [n] \mathbf{d} [n] + \mathbf{w} [n]}_{\mathbf{w}' [n]},         
\end{equation}
with $\mathbf{w}' [n] \in \mathcal{CN}(0, \sigma_{w}^2  \mathbf{I}_{M_{\rm Rx}}  + \sigma_{\rm d}^2 \mathbf{R}_{\rm H} )$ and the channel correlation matrix given by $\mathbf{R}_{\rm H} = \mathbb{E} \NormalBrackets{ \mathbf{H} [n] \NormalBrackets{ \mathbf{H} [n] }^{\rm H} }$.
At the receiver, the channel estimate $\widetilde{\mathbf{H}} [n] \in \mathbb{C}^{M_{\rm Rx} \times M_{\rm Tx}}$ is obtained via \ac{LS} estimation followed by linear interpolation from the quantized received symbols $\mathbf{y}_{\rm q} [n]$.
The quantized symbols are obtained by adding the quantization noise from the \ac{ADC}, following the \ac{AQNM} from~\cite{Dutta2020, Fletcher2007}
\begin{equation}
    \mathbf{y}_{\rm q} [n] =
    \NormalBrackets{ 1 - \Upsilon\NormalBrackets{n_{\rm b}}}  
    \mathbf{y}[n] + 
    \mathbf{w}_{\rm q} [n],
    \label{eq:y_quantized}
\end{equation}
where $\mathbf{w}_{\rm q} [n] \in \mathcal{CN}\NormalBrackets{0,\sigma_{\rm q, Rx}^2 \mathbf{I}_{M_{\rm Rx}}}$ is the quantization noise from  the \ac{ADC} with the power given by
\begin{equation}
    \sigma_{\rm q, Rx}^2 = 
    \Upsilon\NormalBrackets{n_{\rm b}} 
    \NormalBrackets{ 1 - \Upsilon\NormalBrackets{n_{\rm b}}} 
    \mathbb{E} \NormalBrackets{ \norm{\mathbf{y} [n]}_2^2},
\end{equation}
assuming that $\mathbb{E} \NormalBrackets{ \norm{\mathbf{y} [n]}_2^2} = 1 + \sigma_{w}^2$.
To enable optimal beamforming, the estimated channel at the receiver must also be available at the transmitter. 
Since the system operates under a \ac{TDD} protocol, we leverage channel reciprocity to ensure the transmitter has access to the same channel estimate.

\subsubsection{Data Transmission} \label{subsec:data_transmission}
The channel estimate obtained in the training phase is processed according to the advanced channel estimation methods described in~\cite{Pasic2024_TCOM, Lee2016}.
The resulting channel estimate $\overline{\mathbf{H}} [n]$ is then used for precoding and combining.
To obtain optimal performance in terms of \ac{SE}, we apply \ac{SVD} to $\overline{\mathbf{H}} [n]$, with its compact form denoted by
\begin{equation} 
    \overline{\mathbf{H}} [n] = 
    \overline{\mathbf{Q}} [n]  
    \overline{\mathbf{\Sigma}} [n]
    \left( \overline{\mathbf{F}} [n] 
    \right) ^ {\rm H}.
    \label{eq:svd}
\end{equation}
In~\cref{eq:svd}, $\overline{\mathbf{Q}} [n] \in \mathbb{C}^{M_{\rm Rx} \times {\ell_{\rm max}}}$ contains left singular vectors, $\overline{\mathbf{F}} [n] \in \mathbb{C}^{M_{\rm Tx} \times {\ell_{\rm max}}}$ contains right singular vectors, $\overline{\mathbf{\Sigma}} [n]$ is the diagonal matrix of singular values $ \overline{\sigma}_{(1)} [n], \ldots , \overline{\sigma}_{(\ell_{\rm max})} [n]$ and $\ell_{\rm max} = {\rm min} \left( M_{\rm Rx}, M_{\rm Tx} \right)$ denotes
the maximum number of spatial streams.
The power loading matrix $\overline{\mathbf{P}} [n] = {\rm diag}\NormalBrackets{ \overline{p}_{(1)} [n], \ldots , \overline{p}_{(\ell_{\rm max})} [n]} $ is optimized via the water-filling algorithm to maximize the transmission rate, while ensuring compliance with the total transmit power constraint %$ \norm{ \overline{\mathbf{F}}^{\left( \rm m \right)} [n] \left( \overline{\mathbf{P}}^{\left( \rm m \right)} [n] \right)^{1/2}}_F^2 = P_{\rm T}^{\left( \rm m \right)}.$
\begin{equation}
    \norm{ 
    \overline{\mathbf{F}} [n] 
    \left( \overline{\mathbf{P}} [n] \right)^{1/2} 
    }_F^2 = P_{\rm T}.    
    \label{eq:power_normalization}
\end{equation}
The input-output relationship for data transmission is given by
\begin{equation} 
    \mathbf{y} [n] =
    \left( \overline{\mathbf{Q}} [n] \right) ^ {\rm H}
    \mathbf{H} [n] \,
    \overline{\mathbf{x}} [n]
    + \left( \overline{\mathbf{Q}} [n] \right) ^ {\rm H}
    \mathbf{w} [n],        
    \label{eq:data_input_output}
\end{equation}
where $\overline{\mathbf{x}} [n]$ represents the precoded transmit symbols including transmitter impairments $\mathbf{d} [n]$
\begin{equation}
    \overline{\mathbf{x}} [n] = 
    \overline{\mathbf{F}} [n]
    \left( \overline{\mathbf{P}} [n] \right)^{1/2}
    \mathbf{x} [n] 
    + \mathbf{d} [n],
\end{equation}
$\mathbf{x} [n] \in \mathbb{C}^{\ell_{\rm max} \times 1}$ is the transmit symbol vector, $\mathbf{y} [n] \in \mathbb{C}^{\ell_{\rm max} \times 1}$ is the unquantized received symbol vector and $\mathbf{w} [n] \in \mathcal{CN}(0,\sigma_{w}^2 \mathbf{I}_{M_{\rm Rx}})$ is the \ac{AWGN}.
As in~\cref{eq:training_input_output}, we rewrite~\cref{eq:data_input_output} as
\begin{equation} 
	\mathbf{y} [n] =
	\left( \overline{\mathbf{Q}} [n] \right) ^ {\rm H}
	\mathbf{H} [n] \,
	\overline{\mathbf{F}} [n]
	\left( \overline{\mathbf{P}} [n] \right)^{1/2}
	\mathbf{x} [n] 		
	+ \left( \overline{\mathbf{Q}} [n] \right) ^ {\rm H}
	\mathbf{w}' [n].        
	\label{eq:data_input_output_rewritten}
\end{equation}
Finally, the quantized received symbols are given by $\overline{\mathbf{y}}_{\rm q} [n]$, following the approach as in~\cref{eq:y_quantized}.

\subsection{Power Consumption Model} \label{subsec:power_consumption_model}
To account for the power efficiency of the converters, as well as other key components within the \ac{RF} chains that significantly contribute to the overall power consumption, we model the total power consumption of our \ac{MIMO} transceiver as $P_{\rm tot} = P_{\rm Tx} + P_{\rm Rx}$,
where $P_{\rm Tx}$ and $P_{\rm Rx}$ denote the transmitter and receiver power consumptions, respectively.
The transmitter chain consists of \acp{DAC}, mixers and \acp{PA}.
The total transmit power consumption is modeled as~\cite{Dutta2020}
\begin{equation}
    P_{\rm Tx} \, [ \text{mW} ] = M_{\rm Tx} \NormalBrackets{ P_{\rm DC, PA} + P_{\rm LO} + 2 P_{\rm DAC} },
    \label{eq:Tx_power_consumption}
\end{equation}
where $P_{\rm DC, PA}$ is the DC power consumption of the \ac{PA}, $P_{\rm LO}$ is the power consumption of the mixer (driven by the \ac{LO}), and $P_{\rm DAC}$ denotes the power consumed by one \ac{DAC}.
To achieve a desired \ac{EIRP}, the output power per \ac{PA} is given by
\begin{equation}
    P_{\rm out, PA} \, [ \text{dBm} ] = 
    {\rm EIRP} - 20\log_{10} \NormalBrackets{ M_{\rm Tx} }
\end{equation}
and the corresponding DC power drawn by the \ac{PA} is modeled as~\cite{Dutta2020}
\begin{equation}
    P_{\rm DC, PA} \, [ \text{mW} ] = 
    \frac{1}{\eta_{\rm PA}} 
    \NormalBrackets{ 10^{\frac{P_{\rm out, PA}}{10}} - 
    10^{\frac{P_{\rm in, PA} }{10}}},
\end{equation}
where $\eta_{\rm PA}$ is the power added efficiency of the \ac{PA}, and $P_{\rm in, PA}$ is the input power to the \ac{PA} (in dBm). 
The input power to the \ac{PA} is given by
\begin{equation}
    P_{\rm in, PA} \, [ \text{dBm} ] = P_{\rm in, BB} - 10 \log_{10} \NormalBrackets{ M_{\rm Tx} } - {IL}_{\rm mix},
\end{equation}
where $P_{\rm in, BB}$ is the power delivered by the baseband circuitry and ${IL}_{\rm mix}$ is \ac{IL} of the mixer.
The power consumption of the \ac{DAC} is modeled as a function of its resolution $n_{\rm b}$ (in bits) and sampling rate $f_{\rm s}$, and is given by~\cite{Dutta2020}
\begin{equation}
    P_{\rm DAC} \, [ \text{mW} ] = 2^{n_{\rm b}} f_{\rm s} \, {\rm FoM}_{\rm DAC},
\end{equation}
where ${\rm FoM}_{\rm DAC}$ denotes the figure of merit of the \ac{DAC}, typically expressed in femtojoules per conversion step. 

%\textbf{Receiver Power Consumption:}
The receiver chain consists of \acp{LNA}, mixers, and \acp{ADC}.
The total receiver consumption is given by~\cite{Dutta2020}
\begin{equation}
    P_{\rm Rx} \, [ \text{mW} ] = M_{\rm Rx} \NormalBrackets{ P_{\rm DC, LNA} + P_{\rm LO} + 2 P_{\rm ADC} }
    \label{eq:Rx_power_consumption}
\end{equation}
where $P_{\rm DC, LNA}$ is the DC power consumption of the \ac{LNA} and $P_{\rm ADC}$ denotes the power consumed by one \ac{ADC}.
The power consumption of the \ac{ADC} is modeled as a function of its resolution $n_{\rm b}$ (in bits) and sampling rate $f_{\rm s}$, and is given by~\cite{Dutta2020}
\begin{equation}
    P_{\rm ADC} \, [ \text{mW} ] = 2^{n_{\rm b}} f_{\rm s} \, {\rm FoM}_{\rm ADC},
\end{equation}
where ${\rm FoM}_{\rm ADC}$ represents the figure of merit of the \ac{ADC}.

\begin{table}[t]
    \centering
    \caption{Simulation Parameters} 
    \label{tab:simParams}
    \begin{tabular}{rc}
        \hline
        \textbf{Parameter}                          & \textbf{Value} \\ \hline
        Carrier Frequency $f_{\rm c}$               & 25.5\,GHz           \\
        Bandwidth $B$                               & 403.2\,MHz            \\
        Subcarrier Spacing $\bigtriangleup f$       & 120\,kHz             \\
        Sampling Rate $f_{\rm s}$                   & 806.4\,MHz \\
        RF Impairments Power $\sigma_{\rm RF}^2$    & $-$25\,dB~\cite{Roth2018} \\
        Mixer $P_{\rm LO}$ / $IL_{\rm mix}$         & 10\,dBm / 6\,dB~\cite{Gupta2012}  \\
        EIRP / PA Power Added Eff. $\eta_{\rm PA}$  & 30\,dBm / 20\% \\ 
        LNA Power Cons. $P_{\rm DC, LNA}$           & 11\,mW~\cite{Min2007} \\
        Fig. of Merit ${\rm FoM}_{\rm DAC}$/ ${\rm FoM}_{\rm ADC}$ & 65\,fJ/conv / 67.6\,fJ/conv~\cite{Dutta2020} \\
        ADC/DAC Resolution Bits $n_{\rm b}$         & \{2, 4, 8\} \\ \hline  
    \end{tabular}
\end{table}

\begin{figure*}[t]
    \centering
    {\includegraphics[width=2\columnwidth]{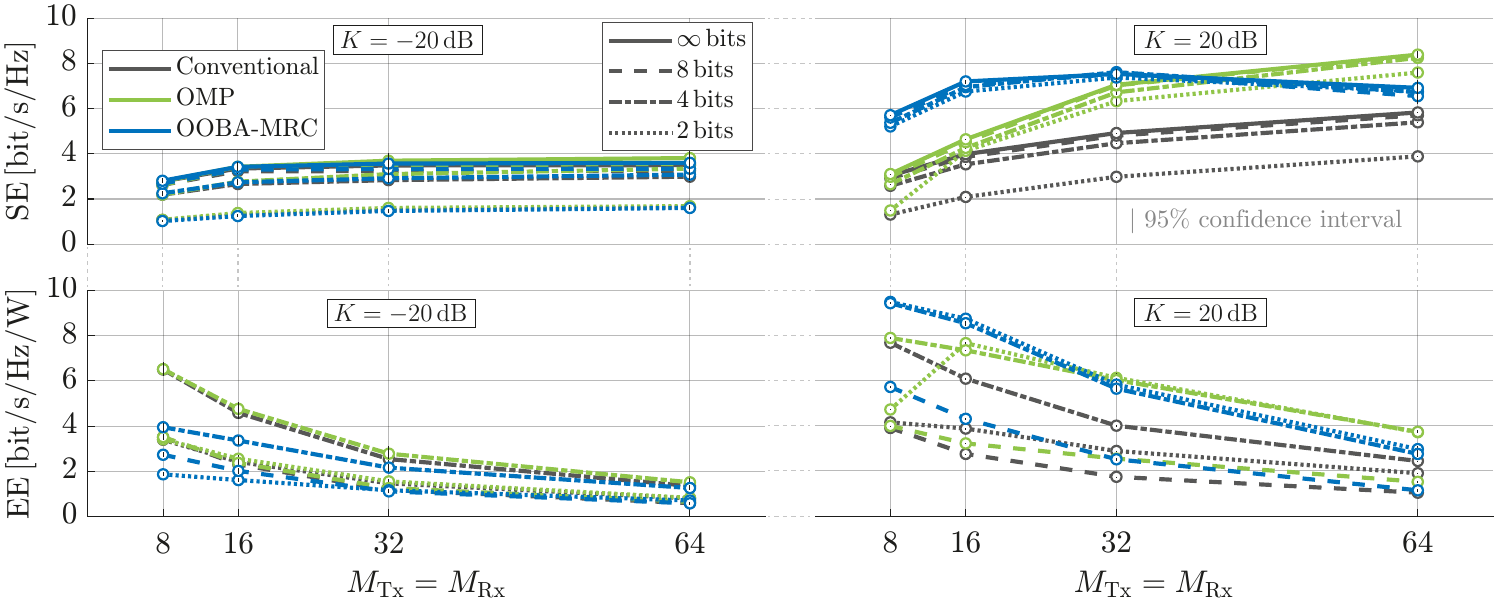}}
    \caption{
    With a low $K$-factor (left), lower resolution reduces \ac{SE} for all methods, while 4-bit quantization yields the highest \ac{EE}.
    At high $K$-factor (right), the \ac{OOBA-MRC} is more robust to quantization, with 2-bit and 4-bit quantizers achieving high \ac{SE} and \ac{EE}.    
    }
    \label{fig:se_ee1}
\end{figure*}

\section{Simulation-based Comparison} \label{sec:comparison}
To evaluate the system performance, we simulate the \ac{SE} and \ac{EE} in a frequency-selective channel.
The transmitter and receiver antenna elements are arranged in a \ac{ULA} configuration with half-wavelength spacing.
The channel model follows the \ac{3GPP} urban macro for the \ac{LOS} scenario with an adjustable $K$-factor (as outlined in~\cite{3gpp.38.901}) and the used simulation parameters are summarized in~\cref{tab:simParams}.
As advanced channel estimation methods, we employ the \ac{OOBA-MRC} method proposed in~\cite{Pasic2024_TCOM} and the \ac{OMP} method detailed in~\cite{Lee2016}.
In addition to the advanced methods, we also include the performance of conventional baseline method, where the channel estimates is set as $\overline{\mathbf{H}} [n] = \widetilde{\mathbf{H}} [n]$.
The achievable \ac{SE} expressed in bits$/$s$/$Hz is defined as
\begin{equation} 
    \mathrm{SE}= 
    \frac{1}{N} 
    \sum_{n=1}^{N} 
    \sum\limits_{\mu=1}^{\ell_{\rm max}} 
    \log_2 \left( 1 + \mathrm{SINR}_{\rm \mu} [n] \right),
    \label{eq:se}
\end{equation}
where the effective \ac{SINR} for the spatial stream ${\rm \mu}$ is given by~\cite{Pasic2024_TCOM}
\begin{equation} 
    \mathrm{SINR}_{\rm \mu} [n] = \frac{ 
    \left| \overline{\mathrm{G}}_{\rm \mu,\mu} [n] \right|^2 }
    { \sum\limits_{\substack{\nu=1 \\ \nu \neq \mu}}^{\ell_{\rm max}} 
    \left| \overline{\mathrm{G}}_{\rm \mu, \nu} [n] \right|^2
    + \sigma_{\rm tot}^2 [n] 
    \norm{ \overline{\mathrm{Q}}_{\rm :,\mu} [n] }^2 },    
    \label{eq:sinr}
\end{equation}
with $\sigma_{\rm tot}^2 [n] =  \sigma_{\rm \mu}^2 [n] +  \sigma_{w}^2 +  \sigma_{\rm d}^2 + \sigma_{\rm q, Rx}^2$.
In~\cref{eq:sinr}, the term $\overline{\mathrm{G}}_{\rm \mu,\nu} [n]$, where $\mu, \nu \in \{ 1, \ldots, \ell_{\rm max} \} $, denotes the entries of the channel gain matrix $\overline{\mathbf{G}} [n] \in \mathbb{C}^{\ell_{\rm max} \times \ell_{\rm max}}$ for subcarrier $n$, defined as
\begin{equation}
    \overline{\mathbf{G}} [n]= 
    \left( \overline{\mathbf{Q}} [n] \right) ^ {\rm H} 
    \mathbf{H} [n]
    \overline{\mathbf{F}} [n] 
    \left( \overline{\mathbf{P}} [n] \right)^{1/2}.
    \label{eq:chgain}
\end{equation}
Furthermore, $ \sigma_{\rm \mu}^2 [n]$ represents the diagonal entries (variances) of the estimation error covariance matrix
$    \overline{\mathbf{C}}_{\varepsilon} [n] = 
    \frac{1}{\ell_{\rm max}}
    \overline{\boldsymbol{\varepsilon}} [n] 
    \left( \overline{\boldsymbol{\varepsilon}} [n] \right)^{\rm H}
$.
The error $\overline{\boldsymbol{\varepsilon}} [n]$ represents the difference between the channel gain matrix computed using the estimated precoder/combiner
$\overline{\mathbf{G}} [n]$, and that obtained using the ideal precoder/combiner $\mathbf{G} [n]$.
The achievable \ac{EE}, expressed in bits$/$s$/$Hz$/$W, is defined by $\mathrm{EE}= \mathrm{SE} / P_{\rm tot}$~\cite{Roth2018}.

We analyze the impact of varying antenna configurations on system performance, with the results in terms of \ac{SE} and \ac{EE} shown in~\cref{fig:se_ee1} for an SNR of 0\,dB.
It can be observed that \ac{SE} mainly increases while \ac{EE} decreases with the number of antennas, regardless of the employed channel estimation method or the considered $K$-factor. 
Furthermore, the achievable \ac{SE} decreases with lower \ac{DAC}/\ac{ADC} resolution.

\textbf{Results at K-factor of $-$20\,dB:}
All methods exhibit comparable performance in terms of \ac{SE}, with the \ac{OMP} method showing slightly improved performance as the number of antennas increases.
The highest \ac{EE} is achieved with 4-bit quantizer resolution across all channel estimation methods.
Notably, the conventional and \ac{OMP} methods exhibit higher \ac{EE} than the \ac{OOBA-MRC} method, which requires the support of the sub-6\,GHz \ac{MIMO} system to achieve comparable \ac{SE}.

\textbf{Results at K-factor of 20\,dB:}
In terms of \ac{SE}, the \ac{OOBA-MRC} method consistently outperforms the conventional method across all \ac{MIMO} configurations, and exceeds the \ac{OMP} method for 8$\times$8, 16$\times$16, and 32$\times$32 configurations.
For 64$\times$64, however, \ac{OMP} achieves higher \ac{SE} due to resolution mismatch between the sub-6\,GHz (8$\times$8) and \ac{MMW} arrays, which degrades \ac{OOBA-MRC} performance.
The conventional method suffers noticeable \ac{SE} loss at lower quantizer resolutions, while \ac{OOBA-MRC} and \ac{OMP} remain more robust. 
In terms of \ac{EE}, the conventional method achieves its best performance with a 4-bit quantizer, while the \ac{OOBA-MRC} and \ac{OMP} methods attain comparable peak performance using either 2-bit and 4-bit quantizers.
The strong \ac{LOS} component enables reliable estimation even at 2-bit resolution. 
Additionally, both advanced methods outperform the conventional method in terms of \ac{EE} across all \ac{MIMO} configurations.

\section{Conclusion} \label{sec:conclusion}
Reducing the quantizer resolution generally leads to a degradation in \ac{SE}.
However, using 4-bit quantizers provides the highest \ac{EE}, thereby offering a favorable trade-off between power consumption and achievable data rate.
At low $K$-factors, all channel estimation methods experience a decline in \ac{SE} as the quantizer resolution is reduced.
At higher $K$-factors, although conventional and \ac{OMP} methods experience performance degradation, the \ac{OOBA-MRC} method remains robust and largely unaffected in terms of \ac{SE}.

% \IEEEtriggeratref{7}
\bibliography{references_short}

@techreport{3gpp.38.901,
 author = {3GPP},
 institution = {{3rd Generation Partnership Project (3GPP)}},
 note = {Version 17.0.0},
 number = {38.901},
 title = {Study on channel model for frequencies from 0.5 to 100 GHz},
 type = {Technical report (TR)},
 year = {2022}
}

@ARTICLE{Pasic2023_mag,
  author={Pasic, Faruk and Di Cicco, Nicola and Skocaj, Marco and Tornatore, Massimo and Schwarz, Stefan and Mecklenbr{\"a}uker, Christoph F. and Degli-Esposti, Vittorio},
  journal={IEEE Communications Magazine}, 
  title={Multi-Band Measurements for Deep Learning-Based Dynamic Channel Prediction and Simulation}, 
  year={2023},
  volume={61},
  number={9},
  pages={98-104},
  doi={10.1109/MCOM.003.2200718}}

@ARTICLE{Liu2020,
  author={Liu, Chunshan and Li, Min and Zhao, Lou and Whiting, Philip and Hanly, Stephen V. and Collings, Iain B.},
  journal={IEEE Transactions on Wireless Communications}, 
  title={{Millimeter-Wave Beam Search With Iterative Deactivation and Beam Shifting}}, 
  year={2020},
  volume={19},
  number={8},
  pages={5117-5131},
  doi={10.1109/TWC.2020.2989343}}

@article{Pasic2024_TCOM,
	author = {Pasic, Faruk and Hofer, Markus and Mussbah, Mariam and Sangodoyin, Seun and Caban, Sebastian and Schwarz, Stefan and Zemen, Thomas and Rupp, Markus and Molisch, Andreas F. and Mecklenbr{\"a}uker, Christoph F.},
	journal = {IEEE Transactions on Communications},
	title = {Millimeter Wave {MIMO} Channel Estimation using sub-6\,{GH}z Out-of-Band Information},
	year = {2024},
        note = {submitted},
        url = {https://owncloud.tuwien.ac.at/index.php/s/3YP9Uq4Dxx17HeZ}}

@ARTICLE{Dutta2020,
  author={Dutta, Sourjya and Barati, C. Nicolas and Ramirez, David and Dhananjay, Aditya and Buckwalter, James F. and Rangan, Sundeep},
  journal={IEEE Transactions on Wireless Communications}, 
  title={{A Case for Digital Beamforming at mmWave}}, 
  year={2020},
  doi={10.1109/TWC.2019.2948329}}

@ARTICLE{Roth2018,
  author={Roth, Kilian and Pirzadeh, Hessam and Swindlehurst, A. Lee and Nossek, Josef A.},
  journal={IEEE Journal of Selected Topics in Signal Processing}, 
  title={A Comparison of Hybrid Beamforming and Digital Beamforming With Low-Resolution {ADCs} for Multiple Users and Imperfect {CSI}}, 
  year={2018},
  doi={10.1109/JSTSP.2018.2813973}}

@ARTICLE{Singh2009,
  author={Singh, Jaspreet and Dabeer, Onkar and Madhow, Upamanyu},
  journal={IEEE Transactions on Communications}, 
  title={On the limits of communication with low-precision analog-to-digital conversion at the receiver}, 
  year={2009},
  volume={57},
  number={12},
  pages={3629-3639},
  doi={10.1109/TCOMM.2009.12.080559}}

@ARTICLE{Fletcher2007,
  author={Fletcher, Alyson K. and Rangan, Sundeep and Goyal, Vivek K and Ramchandran, Kannan},
  journal={IEEE Journal of Selected Topics in Signal Processing}, 
  title={Robust Predictive Quantization: Analysis and Design Via Convex Optimization}, 
  year={2007},
  volume={1},
  number={4},
  pages={618-632},
  doi={10.1109/JSTSP.2007.910622}}

@INPROCEEDINGS{Orhan2015,
  author={Orhan, Oner and Erkip, Elza and Rangan, Sundeep},
  booktitle={2015 Information Theory and Applications Workshop (ITA)}, 
  title={Low power analog-to-digital conversion in millimeter wave systems: Impact of resolution and bandwidth on performance}, 
  year={2015},
  volume={},
  number={},
  pages={191-198},
  doi={10.1109/ITA.2015.7308988}}

@ARTICLE{Molisch2025,
  author={Molisch, Andreas F. and Mecklenbräuker, Christoph F. and Zemen, Thomas and Prokes, Ales and Hofer, Markus and Pasic, Faruk and Hammoud, Hussein},
  journal={IEEE Open Journal of Vehicular Technology}, 
  title={Millimeter-Wave {V2X} Channel Measurements in Urban Environments}, 
  year={2025},
  volume={6},
  number={},
  pages={520-541},
  doi={10.1109/OJVT.2024.3521637}}

@ARTICLE{Lee2016,
  author={Lee, Junho and Gil, Gye-Tae and Lee, Yong H.},
  journal={IEEE Transactions on Communications}, 
  title={Channel Estimation via Orthogonal Matching Pursuit for Hybrid {MIMO} Systems in Millimeter Wave Communications}, 
  year={2016},
  volume={64},
  number={6},
  pages={2370-2386},
  doi={10.1109/TCOMM.2016.2557791}}

@ARTICLE{Gupta2012,
  author={Gupta, Arpit K. and Buckwalter, James F.},
  journal={IEEE Transactions on Microwave Theory and Techniques}, 
  title={Linearity Considerations for Low-{EVM}, Millimeter-Wave Direct-Conversion Modulators}, 
  year={2012},
  doi={10.1109/TMTT.2012.2209435}}

@ARTICLE{Min2007,
  author={Min, Byung-Wook and Rebeiz, Gabriel M.},
  journal={IEEE Microwave and Wireless Components Letters}, 
  title={Ka-Band {SiGe HBT} Low Noise Amplifier Design for Simultaneous Noise and Input Power Matching}, 
  year={2007},
  volume={17},
  number={12},
  pages={891-893},
  doi={10.1109/LMWC.2007.910512}}

@ARTICLE{Ribeiro2018,
  author={Ribeiro, Lucas N. and Schwarz, Stefan and Rupp, Markus and de Almeida, André L. F.},
  journal={IEEE Journal of Selected Topics in Signal Processing}, 
  title={Energy Efficiency of {mmWave} Massive {MIMO} Precoding With Low-Resolution {DACs}}, 
  year={2018},
  volume={12},
  number={2},
  pages={298-312},
  doi={10.1109/JSTSP.2018.2824762}}
\bibliographystyle{IEEEtran}

\end{document}